# Electric field control of labyrinth domain structures in core-shell ferroelectric nanoparticles

Anna N. Morozovska[1, *], Eugene A. Eliseev[2], Salia Cherifi-Hertel[3], Dean R. Evans[4†] and Riccardo Hertel[3‡]

[1] Institute of Physics, National Academy of Sciences of Ukraine,
46, pr. Nauky, 03028 Kyiv, Ukraine

[2] Institute for Problems of Materials Science, National Academy of Sciences of Ukraine,
Krjijanovskogo 3, 03142 Kyiv, Ukraine

[3] Université de Strasbourg, CNRS, Institut de Physique et Chimie des Matériaux de Strasbourg, UMR 7504, 67000 Strasbourg, France

[4] Air Force Research Laboratory, Materials and Manufacturing Directorate,
Wright-Patterson Air Force Base, Ohio, 45433, USA

## Abstract

In the framework of the Landau-Ginzburg-Devonshire (LGD) approach, we studied the possibility of controlling the polarity and chirality of equilibrium domain structures by a homogeneous external electric field in a nanosized ferroelectric core covered with an ultra-thin shell of screening charge. Under certain screening lengths and core sizes, the minimum of the LGD energy, which consists of Landau-Devonshire energy, Ginzburg polarization gradient energy, and electrostatic terms, leads to the spontaneous appearance of stable labyrinth domain structures in the core. The labyrinths evolve from an initial polarization distribution consisting of arbitrarily small randomly oriented nanodomains. The equilibrium labyrinth structure is weakly influenced by details of the initial polarization distribution, such that one can obtain a quasi-continuum of nearly degenerate labyrinth structures, whose number is limited only by the mesh discretization density. Applying a homogeneous electric field to a nanoparticle with labyrinth domains, and subsequently removing it, allows inducing changes to the labyrinth structure, as the maze polarity is controlled by a field projection on the particle polar axis.

---

\*        Corresponding author: anna.n.morozovska@gmail.com

†        Corresponding author: dean.evans@afrl.af.mil

‡        Corresponding author: riccardo.hertel@ipcms.unistra.fr

Under specific conditions of the screening charge relaxation, the quasi-static dielectric susceptibility of the labyrinth structure can be negative, potentially leading to a negative capacitance effect. Considering the general validity of the LGD approach, we expect that an electric field control of labyrinth domains is possible in many spatially-confined nanosized ferroics, which can be potentially interesting for advanced cryptography and modern nanoelectronics.

# I. INTRODUCTION

Since its discovery and until now, nanoscale ferroics (ferromagnets, ferroelectrics, ferroelastics) have been the main object of fundamental research on the physical nature of long-range polar, magnetic, and structural orderings [1, 2, 3]. The emergence of a domain structure, such as polar domains in nanoscale ferroelectrics, and its switching by external electric or/and magnetic fields [4], attracts much attention owing to its relevance for functional nanodevices. In particular, complex topological structures, such as vortices or skyrmions, have been considered as promising functional units for next-generation nanoelectronics. In this context, the interplay between electrostatic effects and epitaxial strain in thin ferroelectric/dielectric multilayer films has been recently exploited to obtain exotic polar textures, which can be electrically-controlled [5]. In contrast to extended thin-film heterostructures, this subject is much less developed in nanoscale ferroelectrics such as core-shell nanoparticles. In this case, the role of surface screening increases significantly with a decrease in ferroic sample size and very often begins to dominate in nanoferroics in comparison with the corresponding bulk contributions [6].

Size and surface screening effects can strongly influence the phase transitions in nanoscale ferroics. They can radically shift the existing phase boundaries and create new metastable states and transient phases of long-range order parameters. For instance, it was predicted theoretically that a decrease of the correlation-gradient polarization energy in the presence of incomplete surface screening can lead to spontaneous bending of uncharged domain walls in ferroelectric thin films [7] and nanoparticles [8]. Such walls can form meandering [7] and/or labyrinth [8] structures. Later, similar structures were observed experimentally in thin ferroelectric films by high-resolution scanning transmission electron microscopy (HR-STEM) [9] and piezoresponse force microscopy (PFM) [10], and the observation was corroborated by *ab initio* calculations [10]. Conductive atomic force microscopy experiments registered enhanced conductivity of curved domain walls in ferroelectrics and multiferroics [11, 12, 13]. Non-Ising and chiral ferroelectric domain walls were predicted theoretically [14] and revealed experimentally by nonlinear optical microscopy [15].

Despite permanent attention being paid to the topic by researchers, there are still many unresolved fundamental and applied problems related to the formation and field control of curved domain structures, such as polar vortices, skyrmions, merons, and labyrinth domains in nanoscale ferroelectrics [10, 16, 17], in which networks can form metastable states and transient phases. Controlling the polarization distribution in metastable states using non-circulating (e.g., homogeneous) electric fields, and tailoring the size and shape of ferroelectric nanoparticles, can open the possibility of creating 3D multi-bit nonvolatile memory with ultra-high density [18, 19].

The problem of creating multilayered nanostructures, where the ferroelectric layer makes a negative contribution to the total capacitance [20], is a promising solution for a significant reduction of power dissipated by field effect transistors and capacitors in chips with an ultra-high density of elements [21, 22, 23]. Note that the total capacitance of an arbitrary system is positive in its stable state, whereas the multiple energy-degenerated metastable states of the ferroelectric domain structure can have a negative capacitance in a quasi-static regime [24].

In an effort to contribute to the solution of the above problems, we study in this paper the possibility of controlling the polarity and chirality of equilibrium labyrinth domain structures in spherical core-shell nanoparticles through the application of an external homogeneous electric field. We consider the case where the core is a uniaxial ferroelectric, and the shell is an ultra-thin semiconducting layer containing mobile screening charges.

## II. CONSIDERED PROBLEM AND SIMULATION DETAILS
### A. Considered Problem and Material Parameters

Let us consider a spherical $Sn_2P_2S_6$ (SPS) nanoparticle of radius $R$ with a ferroelectric polarization $P_3(r)$ directed along z-axis, which corresponds to the crystallographic direction "X" [see **Fig.1(a)**]. The particle is covered by an ultra-thin semiconducting shell, where the mobile screening charge is characterized by the effective surface density $\sigma$ and Debye-Hückel screening length $\lambda$. The screening charges can be localized at surface states [25], which can be induced by strong band-bending via the depolarization field [26, 27, 28]; in this case, $\lambda$ is usually much smaller ($\leq 10$ pm) than one lattice constant (~0.5 nm) [29]. Also, surface charges can be caused by an ionic adsorption at the surface [30, 31], and in this case $\lambda$ can vary from the angstrom to the nm scale. More interestingly can be would be the use of a phase change material that can be thermally tuned from a semiconductor to a metallic state, like e.g., vanadium oxide, and in this case λ values can be tuned from the pm to nm scale in a controllable way. Below, we regard that the nature of the screening shell and its properties determines the value of λ.

A homogeneous electric field directed along z-axis can be applied to the particle and subsequently removed. The field is created by plane electrodes, across which a time-dependent electric voltage, $V(t)$, is applied. Depending on the considered problem, the voltage can be either pulse-like, as shown in **Fig.1(b)**, or sinusoidal, $V(t) = V_{max} sin(\omega t)$.

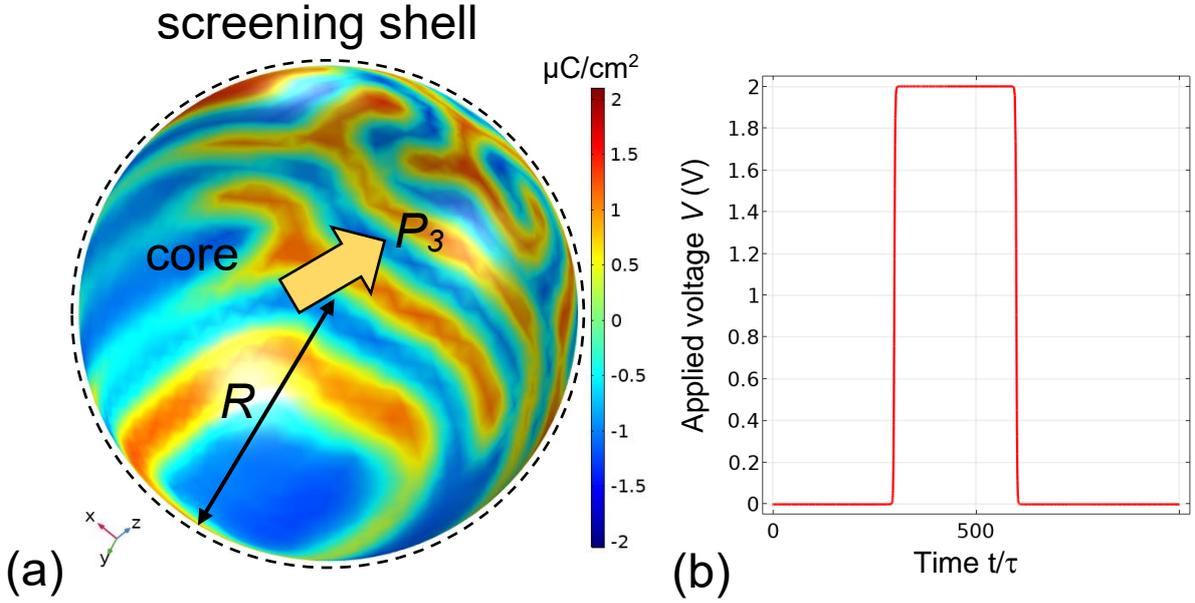

**FIG. 1. A stable labyrinth domain structure in a spherical SPS core covered by an ultra-thin screening shell. (a)** A typical relaxed maze-type (labyrinth) domain state calculated for the core radius $R = 16$ nm, screening length $\lambda = 9$ pm, room temperature 293 K, and time $t \gg 10^3 \tau$, where the parameter $\tau$ is the Landau-Khalatnikov relaxation time of the ferroelectric polarization. SPS parameters are listed in **Table I**. The distance between the electrodes is 44 nm. **(b)** A typical time dependence of the applied voltage used in the calculations.

The ferroelectric polarization $P_3(\mathbf{r})$ contains background and soft-mode contributions, where the corresponding electric displacement $\mathbf{D}$ has the form $\mathbf{D} = \varepsilon_0 \varepsilon_b \mathbf{E} + \mathbf{P}$ inside the particle, and $\mathbf{D} = \varepsilon_0 \varepsilon_e \mathbf{E}$ outside it. Here $\mathbf{E}$ is the electric field, $\varepsilon_b$ is an isotropic background permittivity, and $\varepsilon_e$ is a relative dielectric permittivity of external media. As a rule, $4 \leq \varepsilon_b \leq 10$ [29] and $\varepsilon_e \cong 1$ for air or gaseous environments. The dependence of the in-plane polarization components on $\boldsymbol{E}$ is linear, $P_i = \varepsilon_0(\varepsilon_b - 1)E_i$, where $i = 1, 2$. Hereinafter both designations $x_1 \equiv x$, $x_2 \equiv y$ and $x_3 \equiv z$ are used for the sake of convenience.

The Euler-Lagrange equation for the ferroelectric polarization $P_3(\boldsymbol{r})$ follows from the minimization of the Landau-Ginzburg-Devonshire (LGD) free energy functional,

$$G = G_{L-D} + G_{grad} + G_{el} + G_{es+flexo}, \tag{1a}$$

which includes the Landau-Devonshire energy, $G_{L-D}$, the polarization gradient energy (Ginzburg contribution), $G_{grad}$, the electrostatic contribution $G_{el}$, and the elastic, electrostriction and flexoelectric contributions, $G_{es+flexo}$ [8]:

$$G_{L-D} = \int_{|\vec{r}|<R} d^3r \left( \frac{\alpha}{2}P_3^2 + \frac{\beta}{4}P_3^4 + \frac{\gamma}{6}P_3^6 \right), \tag{1b}$$

$$G_{grad} = \int_{|\vec{r}|<R} d^3r \left( \frac{g_{11}}{2}\left(\frac{\partial P_3}{\partial x_3}\right)^2 + \frac{g_{44}}{2}\left[\left(\frac{\partial P_3}{\partial x_2}\right)^2 + \left(\frac{\partial P_3}{\partial x_1}\right)^2\right] \right), \tag{1c}$$

$$G_{el} = -\int_{|\vec{r}|<R} d^3r \left( P_3E_3 + \frac{\varepsilon_0\varepsilon_b}{2}E_iE_i \right) - \int_{|\vec{r}|=R} d^2r \frac{\phi}{2}\sigma - \frac{\varepsilon_0\varepsilon_e}{2}\int_{|\vec{r}|>R} E_iE_i d^3r, \tag{1d}$$

$$G_{es+flexo} = \int_{|\vec{r}|<R} d^3r \left( -\frac{s_{ijkl}}{2}\sigma_{ij}\sigma_{kl} - Q_{ij3}\sigma_{ij}P_3^2 - \frac{F_{ijk3}}{2}\left(\sigma_{ij}\frac{\partial P_3}{\partial x_k} - P_3\frac{\partial \sigma_{ij}}{\partial x_k}\right) \right). \tag{1e}$$

The coefficient α linearly depends on temperature T, $\alpha = \alpha_T(T - T_C)$, where $T_C$ is the Curie temperature. The coefficients γ, β and the gradient coefficients $g_{11}$, $g_{44}$ are positive and considered to be temperature independent. An isotropic approximation, $g_{44} \approx g_{55}$ in the (001) plane is used for the monoclinic SPS structure. $\sigma_{ij}$ is the stress tensor; $s_{ijkl}$ are elastic compliances; $Q_{ijkl}$ and $F_{ijkl}$ are the electrostriction and flexoelectric tensor components, respectively. The electric field components $E_i$ are derived from the electric potential $\phi$ as $E_i = -\partial\phi/\partial x_i$.

We omit the evident form of the $G_{es+flexo}$ for the sake of simplicity (it is listed in Refs.[32, 33]). Since the values of $F_{ijkl}$ are unknown for SPS, we performed numerical calculations with the coefficients varied in a physically reasonable range, i.e., $|F_{ijkl}| \leq 10^{-11}$ m$^3$/C [34, 35]. The LGD parameters of a bulk ferroelectric SPS are listed in **Table I.**

**Table I.** The parameters for a bulk ferroelectric Sn$_2$P$_2$S$_6$

| Parameter | Dimension | Values for Sn$_2$P$_2$S$_6$ collected from Refs. [36, 37, 38] |
|---|---|---|
| $\varepsilon_b$ | 1 | 7$^*$ |
| $\alpha_T$ | m/F | 1.44×10$^6$ |
| $T_C$ | K | 337 |
| $\beta$ | C$^{-4}$·m$^5$J | 9.40×10$^8$ |
| $\gamma$ | C$^{-6}$·m$^9$J | 5.11×10$^{10}$ |
| $g_{ij}$ | m$^3$/F | $g_{11}$=5.0×10$^{-10}$, $g_{44}$=2.0×10$^{-10**}$ |
| $s_{ij}$ | 1/Pa | $s_{11}$=4.1×10$^{-12}$, $s_{12}$=−1.2×10$^{-12}$, $s_{44}$=5.0×10$^{-12}$ |

| $Q_{ij}$ | m⁴/C² | $Q_{11}$=0.22, $Q_{12}$=0.12, $Q_{12} \approx Q_{13} \approx Q_{23}$ **** |

* estimated from a refraction index value

** the order of magnitude is estimated from the uncharged domain wall width [36-38]

**** estimation of electrostriction is based on thermal expansion data from Say et al. [38].

The corresponding Euler-Lagrange-Khalatnikov equation for $P_3$ has the form:

$$\Gamma \frac{\partial}{\partial t} P_3 + (\alpha - 2Q_{ij3}\sigma_{ij})P_3 + \beta P_3^3 + \gamma P_3^5 - g_{44}\left(\frac{\partial^2}{\partial x_1^2} + \frac{\partial^2}{\partial x_2^2}\right)P_3 - g_{11}\frac{\partial^2 P_3}{\partial x_3^2} = E_3 - F_{ijk3}\frac{\partial \sigma_{ij}}{\partial x_k}$$

(2)

Here $\Gamma$ is the Khalatnikov kinetic coefficient [39], and the corresponding Landau-Khalatnikov relaxation time $\tau$ can be introduced as $\tau = \Gamma/|\alpha|$. The boundary condition for $P_3$ at the spherical surface is "natural", i.e., $\partial P_3/\partial \boldsymbol{n}|_{r=R} = 0$, where $\boldsymbol{n}$ is the outer normal to the surface. The potential $\phi$ satisfies the Poisson equation inside the particle core,

$$\varepsilon_0 \varepsilon_b \Delta \phi = -\frac{\partial P_3}{\partial x_3}, \qquad (3a)$$

and the Laplace equation outside the particle shell,

$$\Delta \phi = 0. \qquad (3b)$$

The 3D Laplace operator is denoted by the symbol $\Delta$. Equations (3) are supplemented by the condition of potential continuity at the particle surface, $(\phi_{ext} - \phi_{int})|_{r=R} = 0$. The boundary condition for the normal components of electric displacements is $\boldsymbol{n}(\boldsymbol{D}_{ext} - \boldsymbol{D}_{int})|_{r=R} = \sigma$, where the surface charge density $\sigma$ is given by the expression $\sigma = -\varepsilon_0 \phi/\lambda$. To model various mobilities of the screening charge, we introduce its relaxation time $\tau_S$, which can be much smaller, the same order of magnitude, or much greater than the Landau-Khalatnikov time $\tau$ [40].

### B. Simulation Details

We perform finite element modeling (**FEM**) in COMSOL@MultiPhysics for different discretization densities of the mesh and polarization relaxation conditions, and observed the spontaneous appearance of stable labyrinth-type domain structures at room temperature from an initial polarization distribution consisting of arbitrarily small randomly oriented nanodomains. The phenomenon takes place under specific screening conditions at the particle surface ($\lambda = 5 - 15$ pm) and for a definite range of particle sizes ($R = 5 - 20$ nm). The origin of the labyrinths' stability is defined by the minimum of the LGD free energy (Eq. 1a). The labyrinth pattern is formed at times $t \gg \tau$. The relaxed pattern is determined by the initial polarization distribution, such that one can obtain a quasi-continuum of energetically almost degenerate labyrinth states; their amount is limited only by the discretization density of the mesh.

The application of a homogeneous electric field to a particle with stable labyrinth domains, followed by its removal, is used to study possible field-induced changes of the labyrinth structure. FEM results, which are presented and analyzed in the next sections, are visualized using Mathematica 12.2 [41].

## III. EFFECT OF SAMPLE HISTORY, MESH PROPERTIES, AND ELECTRIC FIELD ON LABYRINTH DOMAINS

Below we discuss typical FEM results (shown in **Figs. 2-4**) obtained for SPS core-shell nanoparticles at room temperature. Special attention is paid to the effects of arising from the sample history, mesh size, and electric field influence on the labyrinth domains.

The second column (**b**) in **Fig. 2** shows various stable labyrinth domain states developed under "off-field" conditions (at $V = 0$), obtained from different initial polarization distributions, which are shown in the first column (**a**). This result is expected in systems with strong polar anisotropy displaying a quasi-continuum of similar and energy-degenerated states. These labyrinth structures can be called "nominally-equal", because they have basically the same density of domains and domain-wall widths, although microscopically the mazes are not identical. This observation is also consistent with experiments [10, 17].

In the third column (**c**) in **Fig. 2**, which is the same regardless of the different initial distributions, the nanoparticle is completely polarized by an external field. After the removal of the field, column (**d**), the ferroelectric polarization spontaneously relaxes back to the labyrinth structures ($V = 0$). The resulting zero-field patterns are much more similar than before [compare **Fig. 2(b)** with **Fig. 2(d)**]. Relaxed "off-field" stable labyrinth structures, obtained after the system relaxation at $t \gg \tau$, have similar characteristics in terms of the labyrinth patten density, and the position of the domains is very similar in all three cases shown in **Fig. 2(d)**. This similarity is expected given that the "new" initial state (**c**) is nominally identical in all three cases, because the previous configurations are erased from the sample's memory (i.e., sample history) once it is saturated. However, the states in **Fig. 2(d)** are not perfectly identical due to the accumulation of minor numerical errors during the simulation of the polarization evolution from the saturated to the zero-field state.

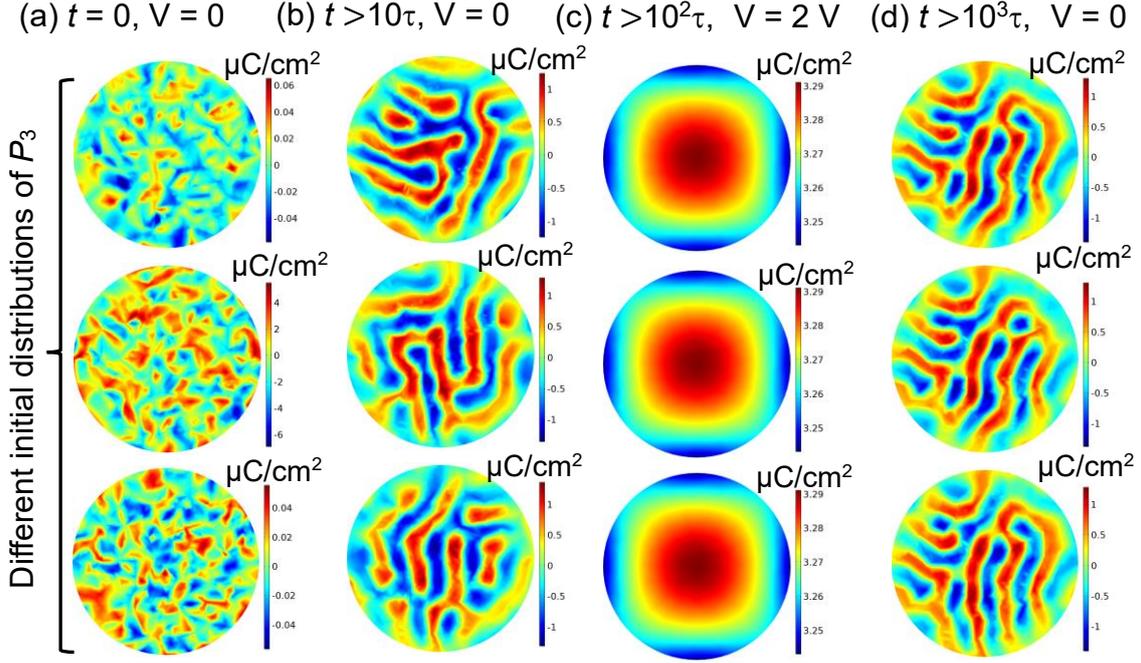

(a) $t = 0$, V = 0     (b) $t > 10\tau$, V = 0     (c) $t > 10^2\tau$,   V = 2 V     (d) $t > 10^3\tau$,   V = 0

Different initial distributions of $P_3$

**FIG. 2. Evolution of the polarization distribution in the equatorial xy-plane of the nanoparticle core calculated for different initial distributions of polarization.** **(a)** Typical initial distributions of polarization consisting of arbitrarily small randomly oriented nanodomains. **(b)** Different labyrinth domain structures, which evolved spontaneously at $V = 0$ from the distributions shown in the column **(a)**. **(c)** A homogeneous external electric field (created by an applied voltage $V = 2$V) destroys the labyrinth domains and induces an almost homogeneous polarization inside the nanoparticle. **(d)** Labyrinth domain structures recover spontaneously after the field is switched off. Other parameters are the same as in **Fig. 1.**

It appears that the stable labyrinth structure depends on the mesh size, but they do not depend on the initial distribution of polarization for the same mesh (see **Fig. 3**). In particular, in **Fig. 3(a)** the mesh discretization density increases (i.e., the mesh's cell average size decreases) from the top to the bottom. Labyrinth structures evolve spontaneously starting from random configurations; the differences in density of the domain walls with increasing discretization density is shown in **Fig. 3(b)** for $t \gg \tau$ at $V = 0$. Next, the on-field saturation of the sample polarization and its off-field relaxation leads to different labyrinth domain states, in this case mesh properties influence the result, as shown in **Figs. 3(c)** and **3(d)**. The samples with low, medium, and high discretization densities show some differences in domain wall width and density. These differences are not significant and almost independent of the sample's history in the case of medium and high mesh discretization densities. The relaxed structure for the finest mesh is noticeably different from those with coarser meshes.

To explain the results shown in **Fig. 3**, two effects should be considered. First, if the physical system displays a quasi-continuum of possible equal-energy states, then it is consistent that different labyrinth structures appear with different meshes. Any small variation, like a change in the mesh density, can result in a different but nominally equal state when starting from a homogeneously polarized state. In a physical experiment, impurities and/or thermal fluctuations could lead to effects similar to those due to a variation of the mesh discretization density in simulations.

The second effect concerns the simulation's accuracy, which is arguably a more delicate aspect. It is known that the accuracy of simulated results depends, among other things, on the mesh's cell size. But this dependence of the numerical error on the cell size is not always gradual and continuous. If the cell size is too large to resolve the changes of the polarization along a domain wall, the errors can be non-negligible. In this case, one cannot be sure that the resulting structure is just a coarser or inaccurate representation of an otherwise correct result. The structure in the top row in **Fig. 3** looks like this could be a borderline case, where the cell size might already be too large. In addition to the possibility of significant discretization errors, there is also an effect of "numerical roughness", which is due to the irregular shape and size of the individual elements in a finite-element mesh. Even when each cell is sufficiently small to resolve continuous changes of the polarization field, spatial variations of the cell size can sometimes affect the results. This numerical effect is to some extent comparable to that of grains, impurities, or other pinning centers in a physical system. Numerically, a region of lower discretization density can represent a preferential site for domain walls and other inhomogeneous structures. This "roughness" effect is usually not crucial if the cell size is small enough, and it becomes negligible as the cell size is further reduced; but changes in the discretization density can sometimes play a role and determine, e.g., whether a domain wall is formed at one position or another if the two positions are otherwise equivalent.

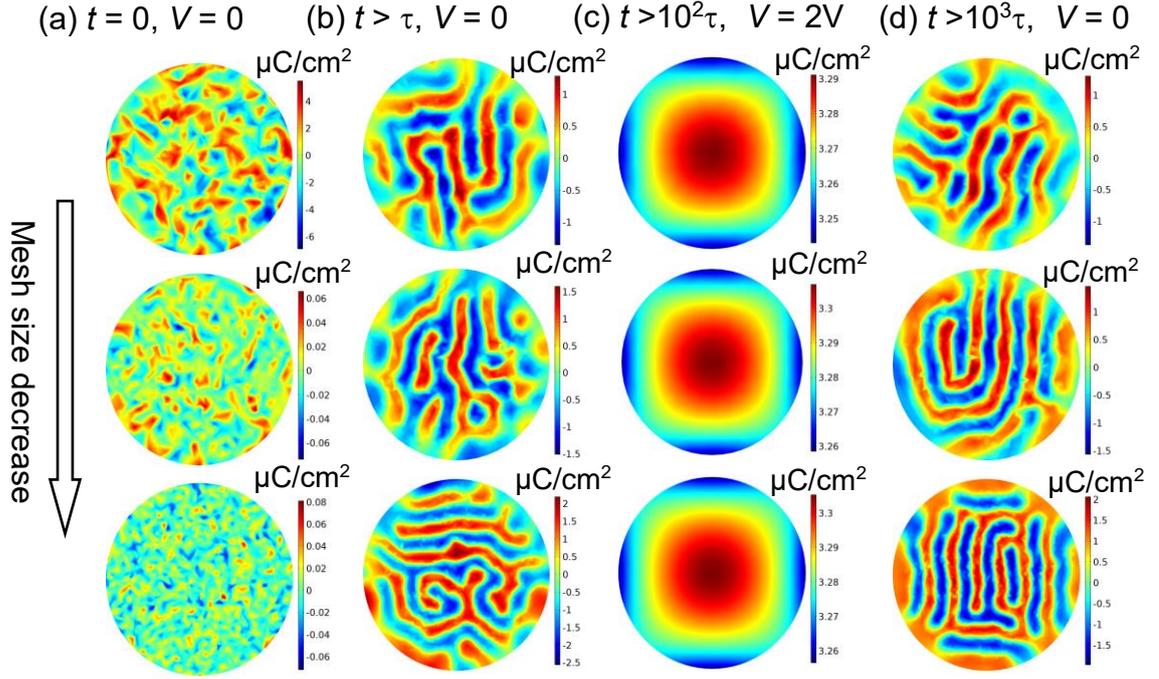

(a) $t = 0$, $V = 0$    (b) $t > \tau$, $V = 0$    (c) $t > 10^2\tau$, $V = 2V$    (d) $t > 10^3\tau$, $V = 0$

Mesh size decrease

**FIG. 3. Evolution of the polarization distribution in the equatorial xy-plane of the nanoparticle core calculated for different mesh cell sizes.** **(a)** Initial distributions of polarization generated using meshes with different discretization densities (the mesh average size decreases from the top to the bottom). **(b)** Different labyrinth domain structures, which evolved spontaneously at $V = 0$ from the initial distributions shown in column **(a)**. **(c)** A homogeneous external electric field (created by $V = 2$ V) destroys the labyrinth domains and induces an almost homogeneous polarization inside the nanoparticle. **(d)** Labyrinth structures recover spontaneously after the field is switched off. Other parameters are the same as in **Fig. 1.**

"Off-field" relaxation of the polarization distribution in the equatorial plane of the nanoparticle, calculated for different initial "on-field" polarization distributions are shown in **Fig. 4**. Top plots **(a)-(e)** are on-field polarization distributions calculated for positive voltages; and bottom plots **(f)-(j)** are on-field polarization distributions calculated for negative voltages. Middle plots show labyrinth domain structures, which recover spontaneously after the positive (plots in the second row) or negative (plots in the third row) voltage is switched off. At voltages higher than the critical value $V_{cr}$ (that is about 1 V for the considered case), the labyrinth polarity is controlled by the field projection on the particle polar axis [compare the plots **(a)** and **(f)**, **(b)** and **(g)**, where red and blue colors interchange]. At voltages lower than the critical value, the external field is not strong enough to erase the previous history of the maze formation. In this case, the field direction can still influence the labyrinth's pattern, and possibly its chirality [compare the plots **(c)** and **(h)**, **(d)** and **(i)**, where the inverted color scheme corresponds to the change in the polarity of the applied field]. However, for voltages below the critical value, the inversion of the color scheme is not complete; it is complete only for the case $|V| > V_{cr}$. Small voltages $|V| \ll V_{cr}$ do not influence the labyrinths in any noticeable way in comparison with off-field relaxation [compare the plots **(e)** and **(j)**].

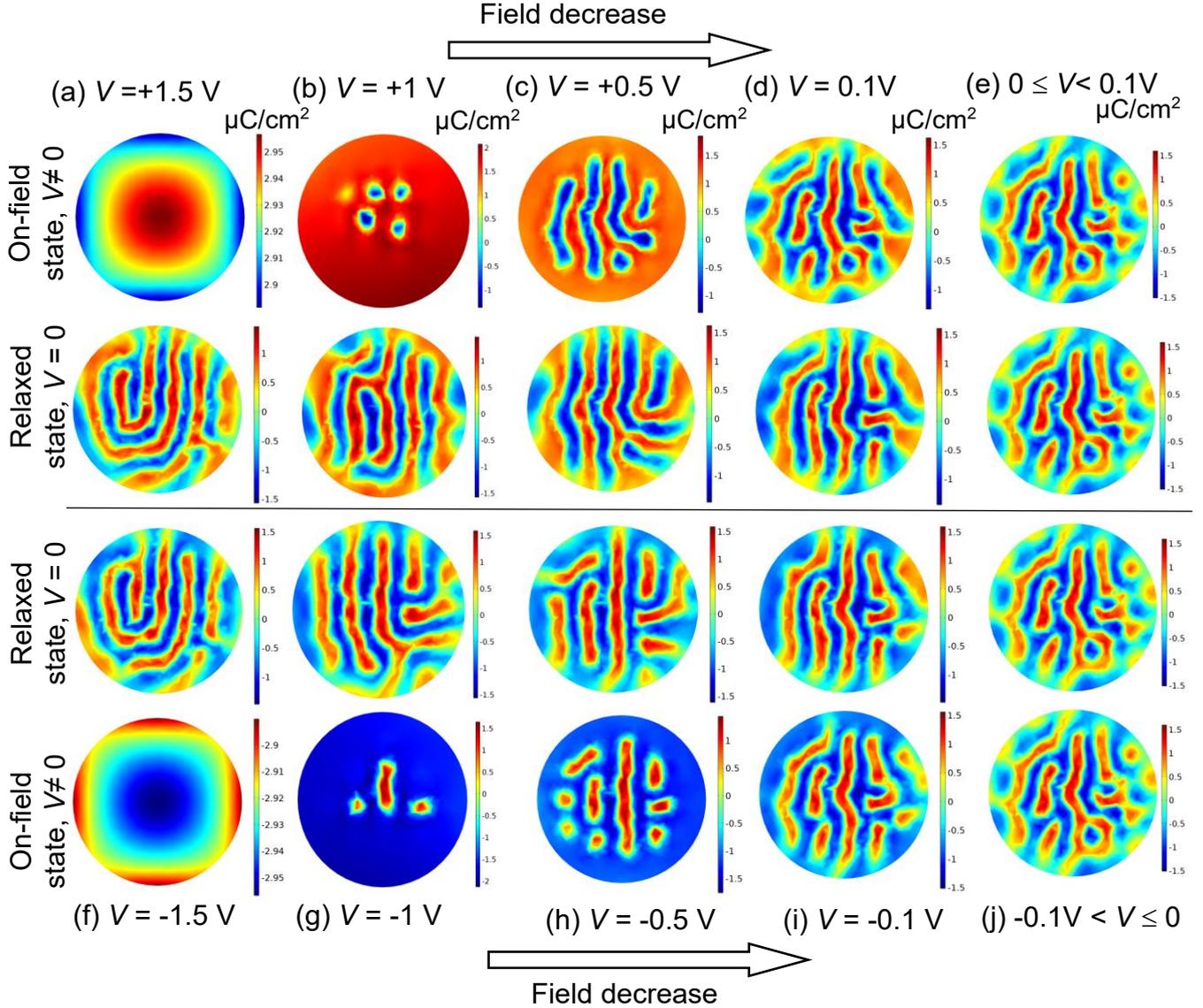

**FIG. 4.** "**Off-field**" **relaxation of the polarization distribution in the equatorial xy-plane of the nanoparticle core, calculated for different initial "on-field" distributions of the polarization.** Top and bottom plots are on-field polarization distributions calculated for different voltages $V$: 1.5 V **(a)**, 1 V **(b)**, 0.5 V **(c)**, 0.1 V **(d)**, 0≤V<0.1 V **(e)**, -1.5 V **(f)**, -1 V **(g)**, -0.5 V **(h)**, -0.1 V **(i)** and **-0.1V** <V≤0 **(k)**. The middle plots show relaxed labyrinth domain structures, which recover spontaneously after the positive (second row) or negative (third row) voltage is switched off. Other parameters are the same as in **Fig. 1.**

## IV. POLARIZATION HYSTERESIS AND EFFECTIVE DIELECTRIC SUSCEPTIBILITY OF LABYRINTH DOMAIN STATES

The orientation of the polarization component $P_3$ can be readily changed by applying an external electric field. Typical hysteresis loops of the polarization $\overline{P_3}(V)$, averaged over the core volume, are shown in **Figs. 5(a)-(c)**. Hysteresis loops of the normalized dielectric susceptibility $\overline{\chi_{33}}(V)$, averaged over the core volume, corresponding to the plots **(a)-(c)**, are shown in **Figs. 5(d)-(f)**. Red parts of the loops are forward runs, blue parts are backward runs, and all initial and transient parts of the curves have been omitted. Every loop is averaged over many $(10-30)$ electric field cycles and possible realizations of mazes. The voltage-symmetric smooth shape of the loops in **Fig. 5** is the result of the averaging.

Quasi-static hysteresis loops of polarization and average dielectric susceptibility exhibit ferroelectric behavior for high amplitude ($V_{max} \gg V_{cr}$, where $V_{cr} \cong 1$V) and an ultra-low frequency ($\omega\tau \leq 10^{-4}$) of the periodic voltage [see **Fig. 5(a)** and **5(d)**]. The polarization loop has a classical rounded square shape, typical for a bulk ferroelectric well below Curie temperature state [see **Fig. 5(a)**]. The labyrinth domain state has been destroyed under the first voltage cycle due to the high external electric field inside the particle. Further cycling of the voltage leads to a quasi-homogeneous polarization reversal at voltages $V = \pm V_c$, where the coercive voltage $V_c \cong 3$ V is relatively high for the considered case. Two sharp maxima of the dielectric susceptibility are located at $V = \pm V_c$, as anticipated for the single-domain ferroelectric state [see **Fig. 5(d)**].

Quasi-static hysteresis loops of polarization and average dielectric susceptibility exhibit antiferroelectric behavior for lower voltage amplitude ($V_{cr} \leq V_{max} \leq 2V_{cr}$) and an ultra-low frequency ($\omega\tau \leq 10^{-4}$) of the applied voltage [see **Fig. 5(b)** and **5(e)**]. Furthermore, the antiferroelectric double shape of the polarization loop is determined by the low-voltage region, $-V_{cr} \leq V \leq V_{cr}$, where the labyrinth domain state is stable and reproducible during every cycle of applied voltage. At $V_{cr} \leq |V| \leq 2V_{cr}$ two very slim loops open due to the labyrinth destruction accompanied by the sharp increase and further saturation of the polarization magnitude [see **Fig. 5(b)**]. The dielectric susceptibility is very small at low voltages and has four sharp maxima located around $V \cong \pm V_{cr}$, as shown in **Fig. 5(e)**.

The hysteresis loops of the polarization exhibit antiferroelectric behavior with two pronounced humps located at voltages slightly greater than $V_{cr}$ for higher, but still quasi-static frequencies ($10^{-3} \leq \omega\tau \leq 10^{-1}$) of the applied voltage [see **Fig. 5(c)**]. The most interesting feature of the humps is the polarization decrease with voltage [see the regions inside dotted ellipses in **Fig. 5(c)**]. Note, the voltage amplitudes within the range $V_{cr} \leq V_{max} \leq 2V_{cr}$ are required for the humps appearance.

It is important to note that the dielectric susceptibility is negative inside the regions of polarization humps [see the regions inside dotted ellipses in **Fig. 5(f)**]. The negative dielectric susceptibility can lead to a quasi-steady state negative capacitance of the core-shell nanoparticle.

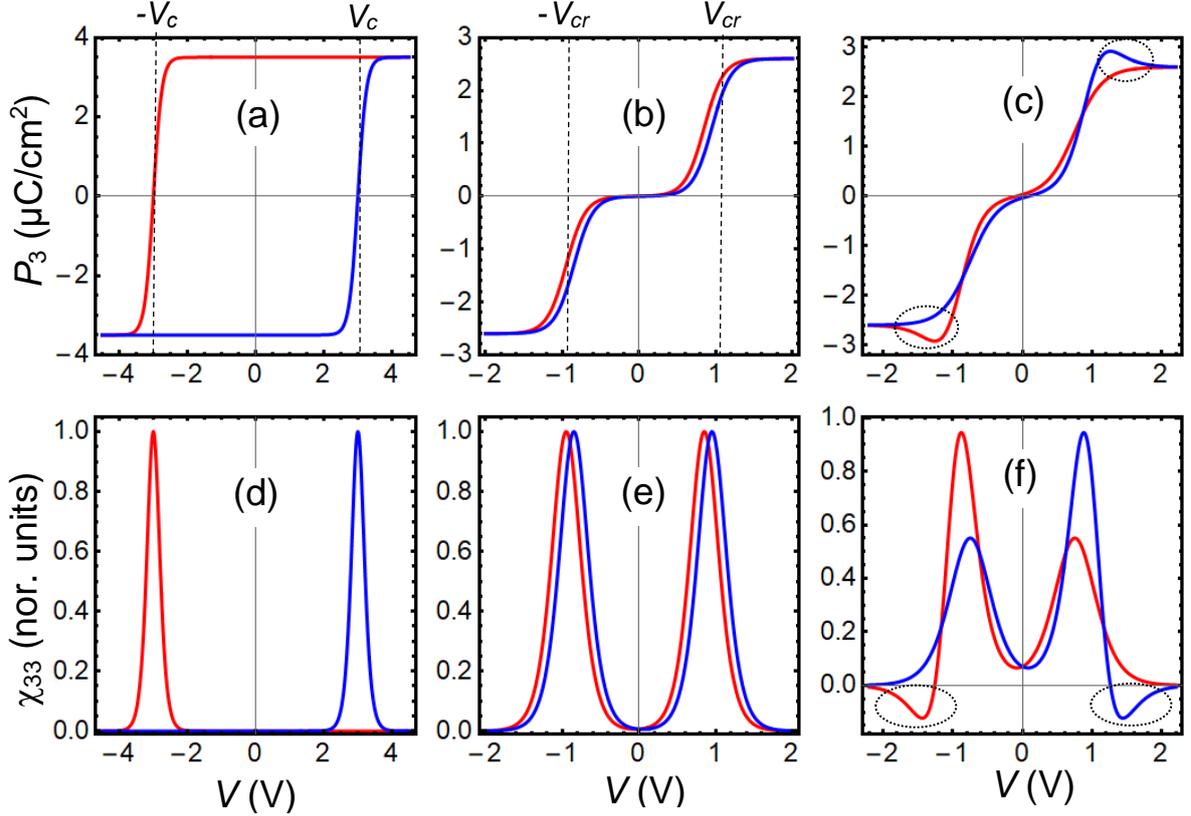

**FIG. 5.** Dependence of the average polarization component $\overline{P_3}$ **(a)-(c)** and the normalized dielectric susceptibility $\overline{\chi_{33}}(V)$ **(d)-(f)** on the applied voltage $V$. Plots **(a)** and **(d)** are calculated for high voltage amplitude ($V_{max}$ =4.5 V) and an ultra-low frequency ($\omega\tau = 10^{-4}$); plots **(b)** and **(e)** are calculated for lower voltage amplitude ($V_{max}$ =2 V) and the same frequency ($\omega\tau = 10^{-4}$); and plots **(c)** and **(f)** are calculated for $V_{max}$ =2 V and a higher frequency ($\omega\tau = 10^{-2}$). Red parts of the loops are forward runs, blue parts are backward runs. Other parameters are the same as in **Fig. 1.**

The physical origin of these anomalous polarization behavior (i.e., humps) and related negative susceptibility are retardation effects in the screening of the depolarization electric fields, which are produced by the labyrinth domain structure in the particle core. The retardation effect occurs in a relatively narrow range of the "favorable" conditions, where these conditions are defined as screening lengths $\lambda = 5 - 15$ pm and relaxation times $\tau < \tau_S < 10\tau$.

A "favorable" frequency range, $10^{-3} \le \omega\tau \le 10^{-1}$, is also required. These conditions mean faster polarization relaxation in comparison with the screening charges, because $\tau < \tau_S$, and a low period of external voltage in comparison with the polarization relaxation time, because $\omega \ll \tau^{-1}$. This also means that the screening cannot be too slow in comparison with external field frequency, since the inequalities $\tau_S < 10\tau$ and $\omega\tau \le 10^{-1}$ leads to $\omega\tau_S \le 1$.

Note that the labyrinth-type and related irregular domain configurations cannot be resolved from macroscopic measurements of the average polarization in a homogeneous electric field. However, they can be reliably observed by local methods using a strong gradient of electric field, such as PFM, which gives information about the distribution of polarization with a nanoscale resolution. Another promising method is resonant elastic soft X-ray scattering, a synchrotron-based method sensitive to chiral polar arrangements through dichroism effects [42, 43].

## V. CONCLUSION

In the framework of the LGD approach, we studied the possibility to control the polarity and chirality of equilibrium labyrinth domain structures in uniaxial ferroelectric core-shell nanoparticles by means of a homogeneous external electric field. Under certain screening conditions at the particle surface and for a definite range of particle sizes, stable labyrinths evolve spontaneously from a random initial polarization distribution with arbitrarily small domains.

The initial polarization distribution determines the details of the relaxed maze pattern. Using different random initial distributions, one can obtain a quasi-continuum of numerous energetically nearly degenerate stable labyrinths, whose number is only limited by a mesh discretization density. These labyrinth structures can be called "nominally-equal", because they have basically the same density of domains and domain-wall widths, although the mazes are different on a microscopic level. The relaxed labyrinth structures with low, medium, and high discretization densities exhibit some differences in domain wall width and density. These differences are not significant and are nearly independent of the sample's history in the case of the meshes with medium and small discretization cell sizes. The relaxed structure for the finest mesh is noticeably different from those obtained with coarser meshes. In a physical experiment, impurities and/or thermal fluctuations could lead to effects similar to those arising from the variation of the discretization density (mesh cell size) in the simulations.

Applying a homogeneous electric field to the particle with labyrinth domains and subsequently removing it allows us to introduce changes in the labyrinth maze. In particular, we find that the maze polarity is controlled by the field projection on the particle polar axis at voltages higher than the critical value. At voltages below the critical value, the external field is not sufficiently strong to erase the previous history of the maze formation, while its direction can still have an influence on the maze's pattern and chirality. In contrast, small voltages do not have an influence on the mazes in any noticeable way in comparison with the off-field relaxation. In view of the general validity of LGD approach, we expect that the electric-field control of labyrinth domains is possible in many spatially-confined ferroics with long-range order parameter, which can be potentially interesting for advanced cryptography.

Under favorable screening conditions, a quasi-continuum of multiple-degenerate labyrinth states correspond to a quasi-static negative dielectric susceptibility. The negative dielectric susceptibility can lead to a quasi-stationary negative capacitance of the core-shell nanoparticles that can be important for their application in advanced nanoelectronics.

**Acknowledgements.** A.N.M. acknowledges EOARD project 9IOE063 and related STCU partner project P751a. E.A.E. acknowledges CNMS2021-B-00843 "Effect of surface ionic screening on polarization reversal scenario in antiferroelectric thin films: analytical theory, machine learning, PFM and cKPFM experiments". R.H. and S.C.H. acknowledge funding from the French National Research Agency through contract ANR-18-CE92-0052 "TOPELEC". S.C.H. acknowledges the Interdisciplinary Thematic Institute EUR QMat (ANR-17-EURE-0024), as part of the ITI 2021-2028 program supported by the IdEx Unistra (ANR-10-IDEX-0002) and SFRI STRAT'US (ANR-20-SFRI-0012) through the French Programme d'Investissement d'Avenir.

# References

1       M. D. Glinchuk, A. V. Ragulya, and V. A. Stephanovich. Nanoferroics. Dordrecht: Springer (2013).

2       J. Hlinka, and P. Ondrejkovic. Skyrmions in ferroelectric materials. Solid State Physics, **70**, 143 (2019), Chapter 4 in "Recent Advances in Topological Ferroics and Their Dynamics" Edited by Robert L. Stamps and Helmut Schulthei. Academic Press (2019).

3       A. Gruverman, M. Alexe, and D. Meier. Piezoresponse force microscopy and nanoferroic phenomena. Nat. Commun. **10,** article number: 1661 (2019).

4       J.-J. Wang, B. Wang, and L.-Q. Chen. Understanding, Predicting, and Designing Ferroelectric Domain Structures and Switching Guided by the Phase-Field Method. Ann. Rev. Mater. Res. **49**, 127 (2019).

5       P. Behera, M. A. May, F. Gómez-Ortiz, S. Susarla, S. Das, C. T. Nelson, L. Caretta, et al. Electric field control of chirality. Science Advances **8**, no. 1, (2022), DOI: 10.1126/sciadv.abj8030.

6       S. V. Kalinin, Y. Kim, D. Fong, and A. N. Morozovska, Surface-screening mechanisms in ferroelectric thin films and their effect on polarization dynamics and domain structures, Rep. Prog. Phys. **81**, 036502 (2018).

7       E. A. Eliseev, A. N. Morozovska, C. T. Nelson, and S. V. Kalinin. Intrinsic structural instabilities of domain walls driven by gradient couplings: meandering anferrodistortive-ferroelectric domain walls in BiFeO₃. Phys. Rev. B **99**, 014112 (2019).

8       E. A. Eliseev, Y. M. Fomichov, S. V. Kalinin, Y. M. Vysochanskii, P. Maksymovich and A. N. Morozovska. Labyrinth domains in ferroelectric nanoparticles: Manifestation of a gradient-induced morphological phase transition. Phys. Rev. B **98**, 054101 (2018).

9       M. J. Han, E. A. Eliseev, A. N. Morozovska, Y. L. Zhu, Y. L. Tang, Y. J. Wang, X. W. Guo, X. L. Ma. Mapping gradient-driven morphological phase transition at the conductive domain walls of strained multiferroic films. Phys. Rev. B **100**, 104109 (2019).

10      Y. Nahas, S. Prokhorenko, J. Fischer, B. Xu, C. Carrétéro, S. Prosandeev, M. Bibes, S. Fusil, B. Dkhil, V. Garcia, and L. Bellaiche, Inverse transition of labyrinth domain patterns in ferroelectric thin films. Nature **577**, 47 (2020).

11      P. Maksymovych, Anna N. Morozovska, Pu Yu, Eugene A. Eliseev, Ying-Hao Chu, Ramamoorthy Ramesh, Arthur P. Baddorf, Sergei V. Kalinin. Tunable metallic conductance in ferroelectric nanodomains. Nano Letters **12**, 209–213 (2012).

12      N. Balke, B. Winchester, W. Ren, Y. H. Chu, A. N. Morozovska, E. A. Eliseev, M. Huijben, R. K. Vasudevan, P. Maksymovych, J. Britson, S. Jesse, I. Kornev, R. Ramesh, L. Bellaiche, L.Q.-Chen, and S.V. Kalinin. Enhanced electric conductivity at ferroelectric vortex cores in BiFeO₃. Nat. Phys. **8**, 81 (2012).

13      R. K. Vasudevan, A. N. Morozovska, E. A. Eliseev, J. Britson, J.-C. Yang, Y.-H. Chu, P. Maksymovych, L. Q. Chen, V. Nagarajan, S. V. Kalinin. Domain wall geometry controls conduction in ferroelectrics. Nano Letters **12**, 5524 (2012).

14      Y. Gu, M. Li, A. N. Morozovska, Y. Wang, E.A. Eliseev, V. Gopalan, and L.-Q. Chen. Non-Ising Character of a Ferroelectric Wall Arises from a Flexoelectric Effect. Phys. Rev. B **89**, 174111 (2014).

15      S. Cherifi-Hertel, H. Bulou, R. Hertel, G. Taupier, K. D. H. Dorkenoo, C. Andreas, J. Guyonnet, I. Gaponenko, K. Gallo, and P. Paruch, Non-ising and chiral ferroelectric domain walls revealed by nonlinear optical microscopy, Nat. Commun. **8**, 15768 (2017).

16       Y. Nahas, S. Prokhorenko, L. Louis, Z. Gui, I. Kornev, and L. Bellaiche. Discovery of stable skyrmionic state in ferroelectric nanocomposites. Nat. Commun. **6**, article no. 8542 (2015).

17      Y. Nahas, S. Prokhorenko, Q. Zhang, V. Govinden, N. Valanoor, and L. Bellaiche. Topology and control of self-assembled domain patterns in low-dimensional ferroelectrics. Nat. Commun. **11**, article no. 5779 (2020).

18      A. N. Morozovska, E. A. Eliseev, R. Hertel, Y. M. Fomichov, V. Tulaidan, V. Yu. Reshetnyak, and D. R. Evans. Electric Field Control of Three-Dimensional Vortex States in Core-Shell Ferroelectric Nanoparticles. Acta Materialia, **200**, 256 (2020).

19      A. N. Morozovska, E. A. Eliseev, Y. M. Fomichov, Y. M. Vysochanskii, V. Yu. Reshetnyak, and D. R. Evans. Controlling the domain structure of ferroelectric nanoparticles using tunable shells. Acta Materialia, **183**, 36 (2020).

20      M. Hoffmann, S. Slesazeck, and T. Mikolajick, Progress and future prospects of negative capacitance electronics: A materials perspective. APL Mater. **9**, 020902 (2021).

21      K. D. Kim, Y. J. Kim, M. H. Park, H. W. Park, Y. J. Kwon, Y. B. Lee, H. J. Kim, T. Moon, Y. H. Lee, S. D. Hyun, B. S. Kim, and C. S. Hwang, Transient negative capacitance effect in atomic-layer-deposited $Al_2O_3/Hf_{0.3}Zr_{0.7}O_2$ bilayer thin film. Adv. Funct. Mater. **29**, 1808228 (2019).

22      A. K. Yadav, et al. Spatially resolved steady-state negative capacitance. Nature **565**, 468 (2019).

23      M. Hoffmann, M. Pešić, K. Chatterjee, A. I. Khan, S. Salahuddin, S. Slesazeck, U. Schroeder, and T. Mikolajick, Direct observation of negative capacitance in polycrystalline ferroelectric $HfO_2$. Adv. Func. Mater. **26**, 8643 (2016).

24      P. Zubko, J. C. Wojdeł, M. Hadjimichael, S. Fernandez-Pena, A. Séné, I. Luk'yanchuk, J.-M. Triscone, and J. Íñiguez, Negative capacitance in multidomain ferroelectric superlattices. Nature **534,** 524 (2016).

25      J. Bardeen, Surface states and rectification at a metal semi-conductor contact. Phys. Rev. **71**, 717 (1947).

26      P. W. M. Blom, R. M. Wolf, J. F. M. Cillessen, and M. P. C. M. Krijn. Ferroelectric Schottky diode. Phys. Rev. Lett. **73**, 2107 (1994).

27      A. N. Morozovska, E. A. Eliseev, S. V. Svechnikov, A. D. Krutov, V. Y. Shur, A. Y. Borisevich, P. Maksymovych, and S. V. Kalinin. Finite size and intrinsic field effect on the polar-active properties of ferroelectric semiconductor heterostructures. Phys. Rev. B **81**, 205308 (2010).

28      Y. A. Genenko, O. Hirsch, and P. Erhart, Surface potential at a ferroelectric grain due to asymmetric screening of depolarization fields. J. Appl. Phys. **115**, 104102 (2014).

29      A. K. Tagantsev and G. Gerra. Interface-induced phenomena in polarization response of ferroelectric thin films. J. Appl. Phys. **100**, 051607 (2006).

30      G. B. StepIehenson and M. J. Highland, Equilibrium and stability of polarization in ultrathin ferroelectric films with ionic surface compensation. Phys. Rev. B, **84**, 064107 (2011).

31      M. J. Highland, T. T. Fister, D. D. Fong, P. H. Fuoss, C. Thompson, J. A. Eastman, S. K. Streiffer, and G. B. Stephenson. Equilibrium polarization of ultrathin $PbTiO_3$ with surface compensation controlled by oxygen partial pressure. Phys. Rev. Lett. **107**, 187602 (2011).

32      I. S. Vorotiahin, E. A. Eliseev, Q. Li, S. V. Kalinin, Y. A. Genenko and A. N. Morozovska. Tuning the Polar States of Ferroelectric Films via Surface Charges and Flexoelectricity. Acta Materialia **137**, 85 (2017).

33      E. A. Eliseev, I. S. Vorotiahin, Y. M. Fomichov, M. D. Glinchuk, S. V. Kalinin, Y. A. Genenko, and A. N. Morozovska. Defect driven flexo-chemical coupling in thin ferroelectric films. Phys. Rev. B **97**, 024102 (2018).

34      "Flexoelectricity in Solids: From Theory to Applications". Ed. by A. K. Tagantsev and P. V. Yudin, World Scientific (2016).

35      B. Wang, Y. Gu, S. Zhang, L.-Q. Chen. Flexoelectricity in solids: Progress, challenges, and perspectives. Progress in Materials Science **106**, 100570 (2019).

36      A. Anema, A. Grabar, and Th. Rasing. The nonlinear optical properties of $Sn_2P_2S_6$. Ferroelectrics **183**, 181 (1996). https://doi.org/10.1080/00150199608224104

37      Yu. Tyagur, Spontaneous Polarization in $Sn_2P_2S_6$ Ferroelectric Single Crystals, Ferroelectrics **345**, 91 (2006).

38      A. Say, O. Mys, A. Grabar, Y. Vysochanskii, and R. Vlokh, Thermal expansion of $Sn_2P_2S_6$ crystals, Phase Transitions **82**, 531 (2009).

39      L. D. Landau, and I. M. Khalatnikov. On the anomalous absorption of sound near a second order phase transition point. In Dokl. Akad. Nauk SSSR **96**, 496 (1954).

40      E. A. Eliseev, M. E. Yelisieiev, S. V. Kalinin, and A. N. Morozovska. Observability of negative capacitance of a ferroelectric film: Theoretical predictions. Phys. Rev. B **105**, 174110 (2022).

41      https://www.wolfram.com/mathematica

42      P. Shafer, P. García-Fernández, P. Aguado-Puente, A. R. Damodaran, A. K. Yadav, C.T. Nelson, S. L. Hsu, J. C. Wojdeł, J. Íñiguez, L. W. Martin, and E. Arenholz, Emergent chirality in the electric polarization texture of titanate superlattices. Proc. Natl. Acad. Sci. U.S.A. **115**, 915 (2018).

43      S. W. Lovesey, G. van der Laan, Resonant x-ray diffraction from chiral electric-polarization structures. Phys. Rev. B **98**, 155410 (2018).